\documentclass[aps,preprint]{revtex4}%
\usepackage{amsfonts}
\usepackage{amsmath}
\usepackage{amssymb}
\usepackage{graphicx}%
\begin{document}
\preprint{ }
\title{Generating conditional atomic entanglement by measuring photon number in a
single output channel}
\author{C. Genes}
\affiliation{Michigan Center for Theoretical Physics, FOCUS Center, and Physics Department,
University of Michigan, Ann Arbor 48109-1040, USA}
\author{P. R. Berman}
\affiliation{Michigan Center for Theoretical Physics, FOCUS Center, and Physics Department,
University of Michigan, Ann Arbor 48109-1040, USA}
\keywords{}
\pacs{42.50. Ct, 42.50. Fx, 42.50. Lc}

\begin{abstract}
The polarization analysis of quantized probe light transmitted through an
atomic ensemble has been used to prepare entangled collective atomic states.
In a "balanced" detection configuration, where the difference signal from two
detection ports is analyzed, the continuous monitoring of a component of the
Stokes field vector provides a means for conditional projective measurements
on the atomic system. Here, we make use of classical driving fields, in the
pulsed regime, and of an "unbalanced" detection setup (single detector) where
the effective photon number of scattered photons is the detected observable.
Conditional atomic spin squeezed states and superpositions of such squeezed
states can be prepared in this manner.

\end{abstract}
\volumeyear{year}
\volumenumber{number}
\issuenumber{number}
\eid{identifier}
\date[Date text]{date}
\received[Received text]{date}

\revised[Revised text]{date}

\accepted[Accepted text]{date}

\published[Published text]{date}

\startpage{1}
\endpage{ }
\maketitle

\section{\bigskip Introduction}

The generation of entanglement in an atomic ensemble of macroscopic dimensions
is a challenging task in the fields of quantum information, quantum
teleportation and precision measurements. Out of the multitude of the
entangled states that could conceivably be engineered using quantum optics
techniques, two subsets have attracted a special interest: spin squeezed
states and "Schrodinger cat" states. The first subset is needed for improving
the precision of quantum measurements beyond the standard quantum limit, while
the second one provides an example of a purely quantum mechanical
superposition at the macroscopic level. The concept of spin squeezing has been
introduced in a comprehensive manner in \cite{spin squeezing concept papers}
which also provide the first theoretical proposals for its realization. Much
attention has been given to the generation of atomic squeezing in the context
of the interaction of cavity quantized fields with atoms, resulting in a
transfer of squeezing to the atoms from a field initially in a squeezed state
\cite{sp. sq. cavity}, or even from an initially coherent field state
\cite{berman genes}. Free space squeezing transfer has also been predicted
\cite{polzik theoretical} and analyzed experimentally \cite{polzik exp.}.

More recent proposals for the generation of entangled atomic states are based
on the dispersive effect of off-resonant light-atom interactions. Information
about the atomic state is correlated with the phase shift accumulated by the
field, and can be "read" by performing a suitable measurement of the field.
The reverse effect of measuring a phase shift acquired by Rydberg atoms in
passing through a microcavity has already been used to generate a "Schrodinger
cat" state of a quantized field in a cavity \cite{haroche}. References
\cite{molmer1, molmer2} provide other examples of phase-shift measurements
that lead to a spin squeezed atomic state and entanglement between two
macroscopic atomic ensembles. It has been shown \cite{kuzmich1} that, with a
suitably chosen internal atomic configuration and quantized fields, for the
off-resonant regime, the interaction can be modelled in the form of a quantum
non-demolition (QND) Hamiltonian proportional to $S_{z}^{a}J_{z}^{f}$, where
$S_{z}^{a}$ represents the signal observable (atomic population spin
component) while $J_{z}^{f}$ (component of the Stokes vector of the field) is
the probe observable to be measured. Most often, a continuous measurement
treatment of the problem is employed, where the polarization analysis of the
probe light transmitted through the medium is detected in a "balanced"
configuration with two photodetectors \cite{takahashi, kuzmich3}. In
\cite{kuzmich2, kuzmich3}, sub-shot-noise fluctuations and spin squeezing with
70\% reduction below the standard quantum limit have been achieved. The
continuous character of the probe monitoring has also rendered possible an
implementation of a feed-back loop for \textit{a posteriori} quantum state
correction based on the conditional measurement of the outcome \cite{mabuchi,
wiseman}.

In this paper a different approach is adopted to analyze the manner in which
entanglement can be generated by means of field detection. In contrast to
other theories, we consider the input field to be a classical field, rather
than a (quantized) coherent state of the field. This is consistent with most
experiments involving large scale entanglement, where the input field is a
laser field, and quantum fluctuations of this field play a negligible role.
The quantum properties of the output field are then related to the radiation
scattered by atoms interacting with the input field.

We apply this treatment to an atomic medium of 4-level Faraday active atoms
($J_{g}=1/2\rightarrow J_{e}=1/2$) that can rotate the direction of
polarization of an incoming field, given a population imbalance between the
ground magnetic sublevels. One prepares the atomic ensemble in a superposition
of ground state spin up and spin down with an equal admixture of both states.
In this case, on average, there is no Faraday effect. Faraday rotation is
induced solely by quantum fluctuations in the population difference of the
spin up and spin down states. With an initially $x$ polarized driving field,
the Faraday effect can be seen as due to a redistribution of photons between
the $x$ and $y$ polarizations of the field. The $y$ polarized part is simply
the source field emitted by the atoms and thus entangled with the atomic
state; in consequence, a measurement of this Faraday rotated part of the field
allows gathering of information on the atomic ensemble with a measurement
strength proportional to the laser power (as shown in \cite{deutsch}).

Our approach also differs from other approaches in that we chose a different
"unravelling" of the dynamics of the system. This is done by the choice of the
measurement basis that we consider, which is given by an "unbalanced"
detection setup with a polarization beam splitter used to separate the $x$
polarized light (directed into an unmonitored port) from the $y$ polarized
part that is directed into a photodetector. Owing to the pulsed regime
considered here, the continuous character of the detection is also lost,
leading to a formulation of the measurement as a "discrete" detection of the
total photon number in the scattered pulse. Conditioned on the outcome of such
a measurement, one can guarantee that the atomic ensemble is in a spin
squeezed state or a coherent superposition of squeezed states ("Schrodinger
cat" state). Only a single operation is necessary here to prepare a bimodal
squeezed state ("Schrodinger cat" state), as opposed to the procedure of
Ref.\cite{massar-polzik} which requires two QND steps to prepare similar cat
states. Intuitively, the reduction from two steps to only one comes from the
indistinguishability of the two possible rotations (left-right) of light
corresponding to a given measured number of scattered photons.

The main results of the paper are obtained in the ideal case where spontaneous
decay to modes other than the optical mode of interest is neglected; the
unavoidable limitations imposed by decay are discussed in Section IV and are
consistent with the general conclusion \cite{molmer1, massar-polzik} that the
resonant optical depth of the atomic ensemble sets the limit on the optimal
achievable entanglement for the collective atomic state. However, we find the
ideal case conclusions to be very useful in terms of providing a clear
physical picture of entanglement generation. Therefore, even if our
measurement strength parameter $C$ is limited by spontaneous decay to values
less than unity, we present results for large values of $C$ to illustrate the
atom-field entanglement. It should be also noted that other limitations will
come into play when a realistic evaluation of the proposed setup is made, such
as non-idealof polarizers, and photodetection dark counts.

The paper is organized as follows. The proposed experimental scheme and some
fundamental notions about spin squeezing are introduced in Section II. Section
III is devoted to the study of the Faraday effect for a single atom and for
many atoms. The single atom scattered field is shown to reflect atomic
population fluctuations, while collective effects in the scattering process
occur for the many atom case. In Sec. IV the effective Hamiltonian needed to
describe the generation of the source field from atomic fluctuations is
derived; the spontaneous emission issue is also addressed here. Also, the
back-action on the atomic state of a measurement of $n$ photons in the field
is derived using a quantum trajectories method. A simple interpretation of the
photon statistics of the scattered field and the connection between
measurement outcomes and projected collective atomic states, is given in Sec.
V. Conclusions are presented in Sec. VI and some detailed calculations for
Sections III and IV are carried out in the Appendices.

\section{Atom-field system}

\begin{figure}
[ptb]
\begin{center}
\includegraphics[
height=2.2961in,
width=4.1658in
]%
{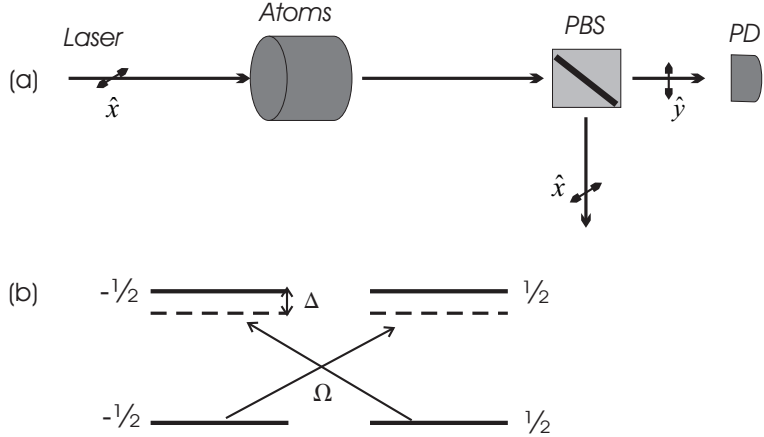}%
\caption{{\footnotesize (a) An }$x${\footnotesize \ polarized classical laser
pulse passes trough a medium of Faraday active atoms, suffering a rotation of
polarization, i.e a }$y${\footnotesize \ polarized field is created. This
scattered field is filtered at the PBS (polarization beam splitter) and
detected at the PD (photodetector). (b) internal structure of an individual
atom in the Faraday active medium.}}%
\label{scheme}%
\end{center}
\end{figure}

An $x$ polarized classical laser pulse (carrier wavelength $\lambda$)
propagating in the $z$ direction [Fig.\ref{scheme}(a)] drives off-resonantly a
$J_{g}=1/2\rightarrow J_{e}=1/2$ \ atomic transition [Fig.\ref{scheme}(b)].
Assuming non-interacting, stationary atoms (number of atoms $N_{a}$ and atomic
density $n_{a}$) in a pencil-shaped medium with transverse area $A$ (matched
to fit the pulse area), length $L$ and Fresnel number close to unity
($F=\frac{A}{\lambda L}\approx1)$, the scattered field intensity is confined
mainly to the forward direction \cite{raymer}. With conveniently chosen
parameters, the resonant optical depth ($d_{res}=n_{a}\lambda^{2}L$) can be
made large, while the off-resonant depth is kept small to ensure a small
saturation of the atomic medium (negligible field depletion). A population
imbalance between the ground state sublevels induces a rotation of the overall
polarization of the field (Faraday effect); in other words, the emerging field
(after the interaction) consists of an unscattered part (whose intensity is
equal to that of the incoming pulse since the depletion of the field due to
absorption is neglected), an $x$ polarized scattered quantized field, and a
$y$ polarized scattered quantized field. The ratio of the $y$ polarized field
amplitude over the incoming field amplitude defines the Faraday rotation
angle. The $y$ component of the emerging field is entangled with the atoms. If
a PBS (polarization beam splitter) is used to filter out the $x$ polarized
field (which is not entangled with the atoms), a measurement at the
photodetector (PD) will generate entanglement within the atomic ensemble.%

Single atom population operators and coherences that are used in what follows
are defined as $\sigma_{g}\left(  m_{g},m_{g}^{\prime}\right)  =\left\vert
g,m_{g}\right\rangle \left\langle g,m_{g}^{\prime}\right\vert $ with
$m_{g},m_{g}^{\prime}=\pm1/2$ (for the ground state manifold), $\sigma
_{e}\left(  m_{e},m_{e}^{\prime}\right)  =\left\vert e,m_{e}\right\rangle
\left\langle e,m_{e}^{\prime}\right\vert $ with $m_{e},m_{e}^{\prime}=\pm1/2$
(excited state manifold) and $\sigma_{+}\left(  m_{e},m_{g}\right)
=\left\vert e,m_{e}\right\rangle \left\langle g,m_{g}\right\vert $ for
$m_{g},m_{e}=\pm1/2$ (raising operator), along with its hermitian conjugate
$\sigma_{-}\left(  m_{e},m_{g}\right)  =\left[  \sigma_{+}\left(  m_{e}%
,m_{g}\right)  \right]  ^{\dagger}$. For the two levels of interest (the
ground magnetic sublevels) collective spin operators are written as sums over
single atom operators%

\begin{align}
S_{z}  &  =%
{\textstyle\sum_{j=1}^{N}}
\left[  \sigma_{g}^{(j)}\left(  \frac{1}{2},\frac{1}{2}\right)  -\sigma
_{g}^{(j)}\left(  -\frac{1}{2},-\frac{1}{2}\right)  \right]  /2\\
S_{+}  &  =%
{\textstyle\sum_{j=1}^{N}}
\sigma_{g}^{(j)}\left(  \frac{1}{2},-\frac{1}{2}\right) \nonumber\\
S_{-}  &  =S_{+}^{\dag}\nonumber\\
S_{x}  &  =\left(  S_{+}+S_{-}\right)  /2\nonumber\\
S_{y}  &  =\left(  S_{+}-S_{-}\right)  /2i\nonumber
\end{align}
The initial state of the collection of atoms is chosen as an eigenstate of
$S_{x}$ with equal populations in the two ground sublevels and initial maximal
built-in coherence. This state can be expanded in eigenstates of $S_{z}$ as%
\begin{equation}
\left\vert S_{x}=S\right\rangle =\overset{S}{\underset{M=-S}{%
{\textstyle\sum}
}}A(S,M)\left\vert S,M\right\rangle \label{initialbinomialstate}%
\end{equation}
Here $A(S,M)=\frac{1}{2^{S}}\sqrt{\frac{(2S)!}{(S+M)!(S-M)!}}$ are binomial
coefficients, while $S=N_{a}/2$. A general squeezing parameter is defined as:%
\begin{equation}
\xi_{\perp}=\sqrt{2S}\Delta S_{\perp}/|\left\langle \mathbf{S}\right\rangle |
\label{sqdef}%
\end{equation}
where $|\left\langle \mathbf{S}\right\rangle |=\sqrt{\left\langle
S_{x}\right\rangle ^{2}+\left\langle S_{y}\right\rangle ^{2}+\left\langle
S_{z}\right\rangle ^{2}}$ is the instantaneous mean spin and $S_{\perp}$ is a
component of the spin orthogonal to the mean spin vector. An atomic system is
said to be spin squeezed when this parameter takes subunitary values. For the
state described in Eq. (\ref{initialbinomialstate}), a squeezed state can be
obtained by minimizing the uncertainty in $S_{z}$ (generating a sub-binomial
distribution), while keeping the average spin in the $x$ direction. A
fundamental quantum limit (Heisenberg limit) for $\xi_{\perp}$ exists and
equals $1/\sqrt{N_{a}}$ \cite{spin squeezing concept papers}.

\section{Faraday effect}

In this section we study the properties of the source field radiated by an
ensemble of four-level atoms [Fig.\ref{scheme}(b)] driven by a classical
field. A single atom treatment is shown to be sufficient to describe the
generation of a Faraday rotation. Some results on the amplitude and intensity
of the source field and on its origin in the quantum mechanical fluctuations
of the atomic population operator are presented here; also the loss of
coherence due to Rayleigh scattering can be analyzed using a perturbative
approach. In a many atom system, the collective radiation effects are
estimated employing a model in which a one-dimensional field is propagating
through a pencil-shaped atomic medium. Maxwell-Bloch calculations are used to
find an expression for the average scattered field amplitude; however, to
study entanglement one requires a mixed semiclassical-quantized approach in
which the driving field-atom interaction is treated semiclassically while the
scattered field is treated quantum mechanically (by taking into account the
interaction of the atoms with the quantum vacuum).

\subsection{One atom}

The positive frequency part of the field (evaluated at position $\mathbf{r}$)
radiated by an atom (at the origin) can be expressed as (see Ref.\cite{loudon})%

\begin{equation}
\mathbf{E}^{(+)}(\mathbf{r},t)\sim\frac{\omega_{0}^{2}}{r}\widehat{\mathbf{r}%
}\times\mathbf{d}^{(-)}(t-\frac{r}{c})\times\widehat{\mathbf{r}}%
\end{equation}
where $\mathbf{d}^{(-)}(t-\frac{\mathbf{r}}{c})$ is the atomic dipole moment
operator at a retarded time. In terms of atomic operators $\mathbf{d}%
^{(-)}(t)=\underset{m_{g},m_{e}}{%
{\textstyle\sum}
}\left\langle g,m_{g}\right\vert \mathbf{d}\left\vert e,m_{e}\right\rangle
\sigma_{-}\left(  m_{e},m_{g};t\right)  $. An average can be performed and the
emitted field is found to be associated with a nonvanishing average atomic
dipole moment. To simplify the calculations we assume that there are two
different time scales; one for the evolution of the coherence between the
ground and excited levels (characteristic time $\gamma^{-1}$), and the other
for ground state populations, which are driven by a pulse with a slowly
varying amplitude (characteristic time $T$ ). If $\gamma T>>1,$ one can solve
the Bloch equations for the coherences and excited state populations at a
given time $t$ taking a 'frozen' value of the Rabi frequency $\chi(t)$ and
ground state populations $\rho_{g,-1/2;g,-1/2}(t)$ and $\rho_{g,1/2;g,1/2}(t)$
at that particular time. A perturbative approach is valid in the limit
$\frac{\chi(t)^{2}}{\Delta^{2}+(\gamma_{e}/2)^{2}}<<1$ for all times $t,$
where $\chi(t)$ is the Rabi frequency, $\Delta$ is the detuning of the laser
field from the resonant atomic frequency and $\gamma$ represents the decay
rate of the upper state manifold \cite{berman}. Expressions for the scattered
field components expectation values (in the incident field direction $z$) in
terms of the radiating ground-excited state coherences are obtained as%

\begin{align}
\left\langle E_{x}^{(+)}(z,t)\right\rangle  &  \sim\left[  \rho_{g,-1/2;e,1/2}%
(t-z/c)+\rho_{g,1/2;e,-1/2}(t-z/c)\right] \\
\left\langle E_{y}^{(+)}(z,t)\right\rangle  &  \sim\left[  \rho_{g,-1/2;e,1/2}%
(t-z/c)-\rho_{g,1/2;e,-1/2}(t-z/c)\right] \nonumber
\end{align}
With adiabatic elimination of the excited state amplitudes, the coherences can
be expressed in terms of the ground state populations only as%

\begin{align}
\rho_{g,-1/2;e,1/2}(t)  &  =\frac{-\chi(t)/\sqrt{3}}{\Delta-i\frac{\gamma}{2}%
}\rho_{g,-1/2;g,-1/2}(t)\\
\rho_{g,1/2;e,-1/2}(t)  &  =\frac{-\chi(t)/\sqrt{3}}{\Delta-i\frac{\gamma}{2}%
}\rho_{g,1/2;g,1/2}(t)\nonumber
\end{align}
and the fields become%

\begin{align}
\left\langle E_{x}^{(+)}(z,t)\right\rangle  &  \sim\chi(t)\left\{
\rho_{g,1/2;g,1/2}(t-z/c)+\rho_{g,-1/2;g,-1/2}(t-z/c)\right\} \\
\left\langle E_{y}^{(+)}(z,t)\right\rangle  &  \sim\chi(t)\left\{
\rho_{g,1/2;g,1/2}(t-z/c)-\rho_{g,-1/2;g,-1/2}(t-z/c)\right\} \nonumber
\end{align}
The $y$ part is the Faraday rotated field and is seen to be produced (at least
on average) by the population imbalance between the ground substates. However,
the scattered field intensity (for both $x$ and $y$ polarized scattered
fields) is isotropic and reflects only the variation of the driving field
envelope $I_{x,y}(z,t)\sim\left\langle E_{x,y}^{(-)}(z,t)E_{x,y}%
^{(+)}(z,t)\right\rangle \sim\left\vert \chi(t)\right\vert ^{2}$. We conclude
that, when one measures the ratio of field amplitude $\left\langle E_{y}%
^{(+)}(z,t)\right\rangle $ to the driving field amplitude, the noise affecting
its detection (given by the ratio of $I_{y}(z,t)$ to $\left\vert
\chi(t)\right\vert ^{2}$) is always larger than the maximum signal. This
problem can be overcome by making use of an ensemble of atoms where the signal
scales as $N_{a}^{2}$ while the noise scales only as $N_{a}$.

A final observation that we make and which will be useful when the decoherence
effects of the spontaneous decay will be taken into account is that the ground
state coherence decays at a rate proportional to the driving field intensity.
The parameter describing the total loss of coherence during the time the laser
pulse is on is the time integrated one atom spontaneous emission rate:%

\begin{equation}
C_{spon}^{2}=\gamma\overset{T}{\underset{0}{%
{\textstyle\int}
}}dt\frac{\left\vert \chi(t)\right\vert ^{2}}{\Delta^{2}} \label{Cspon}%
\end{equation}

\subsection{Maxwell-Bloch approach}

Having observed that the polarization rotation can be explained as a single
atom effect, we calculate the response of a collection of such Faraday active
atoms to an incident classical field. A pencil-shaped medium (length $L$) with
transverse area $A$ and density $n=N_{a}/AL$ is considered. Solving the Bloch
equations for the atomic polarizability induced by the external field and
substituting this back into the Maxwell equations, we obtain the induced
amplitude and phase changes of the original field. The main result of Appendix
A is that the Faraday rotation is given by%

\begin{equation}
\phi=\frac{n\left\vert p\right\vert ^{2}\Omega L}{\hbar\Delta c\epsilon_{0}%
}[\rho_{g,1/2;g,1/2}(t)-\rho_{g,-1/2;g,-1/2}(t)]=\frac{2p^{2}\Omega}%
{\hbar\Delta c\epsilon_{0}A}\left\langle S_{z}\right\rangle \label{Max-Bloch}%
\end{equation}
and is proportional to the average population imbalance $\left\langle
S_{z}\right\rangle $ of the entire ensemble. However, this analysis provides
an expression for the average effect only. Notice that for a whole class of
initial states exhibiting different population fluctuations but with the same
vanishing expectation value of $S_{z}$, a vanishing signal is obtained. This
comes from ignoring the interaction with the vacuum field or, equivalently, by
treating the scattered field as classical.

\subsection{Source field approach}

When the vacuum modes are taken into account, a mixed semiclassical-quantized
field approach can be used in which the atomic dipole is driven by a classical
field and radiates a source field in the quantized vacuum. The source field
components in the forward direction at position $z$ are obtained as (see
Appendix B)%

\begin{align}
E_{x}^{(+)}(z,t)  &  =\frac{-inL\Omega\left\vert p\right\vert ^{2}}%
{2\hbar\Delta\varepsilon_{0}c}E_{0}(0,t-z/c)e^{-i\Omega(t-z/c)}\\
E_{y}^{(+)}(z,t)  &  =\frac{nL\Omega\left\vert p\right\vert ^{2}}{2\hbar
\Delta\varepsilon_{0}c}E_{0}(0,t-z/c)e^{-i\Omega(t-z/c)}\frac{S_{z}}%
{S}\nonumber
\end{align}
The $y$ part (unlike the $x$ part) shows a dependence on a collective spin
operator of the ensemble. A Faraday rotation operator can be defined as%
\begin{equation}
\widehat{\phi}=2E_{y}^{(+)}(z,t)e^{i\Omega(t-z/c)}/E_{0}(z/c,t)=\frac
{nL\Omega\left\vert p\right\vert ^{2}}{\hbar\Delta\varepsilon_{0}c}\frac
{S_{z}}{S}%
\end{equation}
having variance%

\begin{equation}
\Delta\widehat{\phi}=\frac{n\left\vert p\right\vert ^{2}\Omega L}{\hbar\Delta
c\varepsilon_{0}}\frac{\Delta S_{z}}{S}%
\end{equation}
The mean value of the rotation angle is, as expected, the same as predicted in
the previous subsection. The usefulness of the new result is that atomic
states that exhibit different fluctuations give rise to source fields with
different quantum properties. The detection of such a field can therefore
provide information about collective atomic fluctuations in the ensemble.

\section{Effective Interaction Hamiltonian}

A simple form for the evolution generator of the system atoms-forward
scattered field can be obtained by neglecting the coupling of atoms to any
other field modes. In other words, spontaneous emission is neglected and the
system follows a deterministic evolution described by an effective Hamiltonian
that is derived below. Spontaneous emission can be introduced
phenomenologically and will produce a degradation of the measurement-induced entanglement.

\subsection{Effective Hamiltonian in the absence of decay}

As seen in the previous section, the field operator (in the Heisenberg
picture) for the $y$ polarized mode is proportional to the atomic operator
$S_{z}$. That suggests that entanglement between the field and atoms is
present; however, a wave function (or density matrix) approach is needed to
quantitatively estimate the extent of this entanglement. Reducing this
situation to the problem of a driven quantized dipole radiating into the
vacuum modes, one can find an effective interaction Hamiltonian (for
derivation, see Appendix C). With the observation (justified in Appendix C)
that the evolution of the $x$ polarized mode is decoupled from the evolution
of the atomic state and of the $y$ polarized mode, the evolution operator
generated by the Hamiltonian is%

\begin{equation}
U(T)=e^{-iC(c_{y}^{\dag}-c_{y})S_{z}},
\end{equation}
where the interaction parameter $C=\left[  \frac{3}{16\pi^{2}}\left(
\frac{\lambda^{2}}{A}\right)  \left(  \gamma\overset{T}{\underset{0}{%
{\textstyle\int}
}}dt\frac{\left\vert \chi(t)\right\vert ^{2}}{\Delta^{2}}\right)  \right]
^{1/2}$. The operator $c_{y}$ is the annihilation operator for the source
field, that, when applied to the vacuum, creates a quasimonochromatic $y$
polarized one photon pulse of energy $\hbar\Omega$ and duration $T$,
propagating in the positive $z$ direction.

The state vector for the atom-field system is given by%
\begin{align}
\left\vert \Psi(T)\right\rangle  &  =U(T)\left\vert S_{x}=S\right\rangle
\otimes\left\vert 0\right\rangle _{y}=\overset{S}{\underset{M=-S}{\sum}%
}A(S,M)\exp[-iC(c_{y}^{\dag}-c_{y})S_{z}]\left\vert 0\right\rangle
_{y}\left\vert S,M\right\rangle =\label{statevectorafterinteraction}\\
&  =\overset{S}{\underset{M=-S}{\sum}}A(S,M)\exp[-iC(c_{y}^{\dag}%
-c_{y})M]\left\vert 0\right\rangle _{y}\left\vert S,M\right\rangle \nonumber\\
&  =\overset{S}{\underset{M=-S}{\sum}}A(S,M)\left(  \left\vert
-iCM\right\rangle _{y}^{coh}\right)  \otimes\left\vert S,M\right\rangle
=\nonumber\\
&  =\overset{S}{\underset{M=-S}{\sum}}A(S,M)e^{-(CM)^{2}/2}\left(
\overset{\infty}{\underset{n_{y}=0}{\sum}}\frac{(-iCM)^{n_{y}}}{\sqrt{n_{y}!}%
}\left\vert n_{y}\right\rangle _{y}\right)  \otimes\left\vert S,M\right\rangle
\nonumber
\end{align}

The evolution generator is a displacement operator whose amplitude is an
operator rather than the usual c-number. As such it corresponds to a
superposition of coherent states, each of which is associated with one of the
eigenstates of $S_{z}.$ It is evident from
Eq.(\ref{statevectorafterinteraction}) that the participation of each atomic
state to the final amplitude of the scattered field is governed by the
distribution of the atomic fluctuations in the initial atomic state, namely
the binomial coefficient $A(S,M)$. It is also seen that the detection of a
number of photons in the proximity of one of the values $C^{2}M^{2}$ indicates
that one of the two atomic state with projections $\pm M$ is responsible for
scattering. However, since no phase information on the measured field is
available, the states are indistinguishable, leading to a collapse of the
atomic state onto a 'Schrodinger cat' state. An exception is the state with
$M=0$, which is simply connected with an outcome of zero scattered photons;
such a 'null measurement' leads to the squeezing of the initial binomial distribution.

\subsection{Inclusion of spontaneous emission}

In the above derivation, the coupling of atoms to field modes other than those
belonging to the optical mode of interest (forward scattered field) is
neglected. In consequence, even if the theory correctly describes the
measurement strength that applies to a detection of the source field, it fails
to account for the decay of the collective atomic coherence resulting from
spontaneous emission into other vacuum modes. The coherence loss associated
with $C$ is only a fraction, $\lambda^{2}/$ $A$, of the exact loss given by
Eq. (\ref{Cspon}). The interpretation is straightforward: in looking at the
coupling to the forward scattered field only spontaneously emitted photons in
a solid angle $\lambda^{2}/$ $A$ are considered.

The effect of spontaneous decay can be taken into account in two ways: either
by limiting the number of spontaneously emitted photons per atom to a
negligible value and proceed with an estimate of the optimum achievable
entanglement, or by including the correct decay rate ($C_{spon}^{2}$) in the
expression for the evolution of the total mean spin of the sample (while the
measurement process will still be described by a strength $C^{2}$). The first
approach (see Refs. \cite{massar-polzik, polcirac}), discussed in the
following, requires a constant total spin of the sample (on the surface of the
Bloch sphere), while in the second approach the optimization of the squeezing
parameter as a function of $C$ will provide the exact limitation (section VI).

For a pulse consisting of $N_{ph}$ photons, off-resonantly interacting with
the ensemble of atoms under consideration, the total loss is given by
$d_{off-res}N_{ph}$, where $d_{off-res}$ is the off-resonance optical depth of
the sample. The condition of less than one photon loss per atom reads
$\eta=d_{off-res}N_{ph}/N_{a}<1$ which leads to $\eta=\left(  \frac{d_{res}%
}{N_{a}}\right)  \left(  \frac{\gamma}{\Delta}\right)  ^{2}N_{ph}<1$ .
Expressing $C$ in terms of the number of incoming photons $C\simeq\left(
\frac{\gamma}{\Delta}\right)  \left[  \frac{\lambda^{2}}{A}\right]
\sqrt{N_{ph}}=\frac{\gamma}{\Delta}\left(  \frac{d_{res}}{N_{a}}\right)
\sqrt{N_{ph}}$, one can write $C=$ $\eta^{1/2}\sqrt{\frac{d_{res}}{N_{a}}}$
from which it is deduced that%
\begin{equation}
C<\sqrt{\frac{d_{res}}{N_{a}}} \label{Climit}%
\end{equation}

As a consequence, $C$ is always less than unity; nevertheless, spin squeezing
of order $\sqrt{1/d_{res}}$ can still be achieved. For the regime in which the
above condition is satisfied the effect of emission in directions other than
into the mode subjected to detection can be safely omitted. The length of the
mean spin stays approximately constant; in view of Eq. (\ref{sqdef}), the
evolution of the orthogonal spin component variance as a result of the
measurement is enough to describe squeezing. However, in the remaining
sections, we'll use the second approach where the evolution of the system
described by the effective Hamiltonian for any value of the measurement
strength parameter is corrected by a fast decoherence of the average spin due
to the inclusion of spontaneous decay.

\section{Measurement Process}

An indirect detection scheme can be imagined that closely resembles the case
of the monitoring of a decaying field in a cavity \cite{cavity field
monitoring}. The system is represented by the source field plus atoms, while
the photodetector is the environment. The monitoring of the photons that
escape to the environment is made through the detection of photoelectrons. It
is assumed that every absorbed photon produces one photoelectron which is
registered with 100\% efficiency. The detection process lasts for a time $T$,
which is the time the source field interacts with the photodetector. The field
passing through the detector is attenuated at a rate $\lambda$ which gives a
total attenuation about $e^{-\lambda T}$ of the field during detection. With
sufficiently large $\lambda$, all source field photons are detected, a
condition that is equivalent to a 100\% detection efficiency.

The formalism we use is one of continuous measurement theory. Our system of
field and atoms loses photons at a rate $\lambda$. We then view the process as
a piecewise deterministic process where periods of deterministic evolution are
interrupted by sudden quantum jumps induced by the detection of a photon. The
Lindblad jump operator is the one-mode effective annihilation operator for the
source field $c$. The free evolution is determined by the nonhermitian
operator: $H_{nh}=-i\hbar\frac{\lambda}{2}c^{\dag}c$ which generates an
evolution operator $U_{d}(t)=e^{-\frac{\lambda t}{2}c^{\dag}c}.$ We are
interested in the state of field and atoms after a detection time $T_{d}$ ;
therefore with notation $1-e^{-\lambda t}=\mu$, the evolution operator becomes
$U_{d}(T_{d})=(1-\mu)^{c^{\dag}c}.$

The final state averaged over all detection histories (one history is a
sequence of detection times inside the detection interval) that lead to a
number of $n_{m}$ detected photons (up to a phase factor) is%

\begin{equation}
\left\vert \Psi_{n_{m}}\right\rangle =\frac{c^{n_{m}}U_{d}(T_{d})\left\vert
\Psi_{initial}\right\rangle }{\left\vert \left\vert c^{n_{m}}U_{d}%
(T_{d})\left\vert \Psi_{initial}\right\rangle \right\vert \right\vert }%
\end{equation}
From Eq. (\ref{statevectorafterinteraction}), we obtain%

\begin{equation}
\left\vert \Psi_{n_{m}}\right\rangle _{F+A}=\overset{S}{\underset{M=-S}{%
{\textstyle\sum}
}}\frac{A(S,M)(iCM)^{n_{m}}e^{-\mu(CM)^{2}/2}}{\sqrt{\overset{S}%
{\underset{X=-S}{\sum}}\left\vert A(S,M)\right\vert ^{2}(CM)^{2n_{m}}%
e^{-\mu(CM)^{2}}}}]\left\vert S,M\right\rangle \otimes\left\vert
iCM(1-\mu)\right\rangle _{y}^{coh} \label{imperfectdetection}%
\end{equation}
Two expected results are transparent here. Firstly, a field in a coherent
state undergoing photodetection is left in a smaller (attenuated) amplitude
coherent state. Secondly, when the condition of 100\% detection efficiency is
imposed by setting $\mu=1,$ the state of the field after measurement is the
vacuum field. The decoupled entangled final state for the atoms is then given by%

\begin{equation}
\left\vert \Psi_{n_{m}}\right\rangle _{A}=\overset{S}{\underset{M=-S}{%
{\textstyle\sum}
}}\frac{A(S,M)(iCM)^{n_{m}}e^{-(CM)^{2}/2}}{\sqrt{\overset{S}{\underset
{X=-S}{\sum}}\left\vert A(S,M)\right\vert ^{2}(CM)^{2n_{m}}e^{-(CM)^{2}}}%
}]\left\vert S,M\right\rangle \label{statevectorafterdetection}%
\end{equation}

\section{Generated Atomic Entanglement}

During the detection window (which is long enough to include the whole
scattered field pulse) the detector registers a number of clicks which we
regard as the outcome of the measurement. An outcome of $n_{m}$ clicks
indicates a particular set of atomic states that are most likely to have given
rise to a field with $n_{m}$ photons. Consequently, owing to the entanglement
generated by the interaction, in response to the performed measurement, the
atomic system is projected onto this set of states with properties that are
analyzed below.

\subsection{Photon statistics}

Equation (\ref{statevectorafterinteraction}) describes the field as a
collection of coherent states with different amplitudes, each of them
proportional to the corresponding atomic state that is responsible for its
scattering. After tracing over the atomic states, a photon number distribution
can be obtained, given by%

\begin{equation}
P_{n_{m}}(C,S)=\overset{S}{\underset{N=-S}{%
{\textstyle\sum}
}}\left\vert A(S,N)\right\vert ^{2}e^{-(CN)^{2}}\frac{(CN)^{2n_{m}}}{n_{m}!}%
\end{equation}

\begin{figure}
[ptb]
\begin{center}
\includegraphics[
height=2.476in,
width=4.1451in
]%
{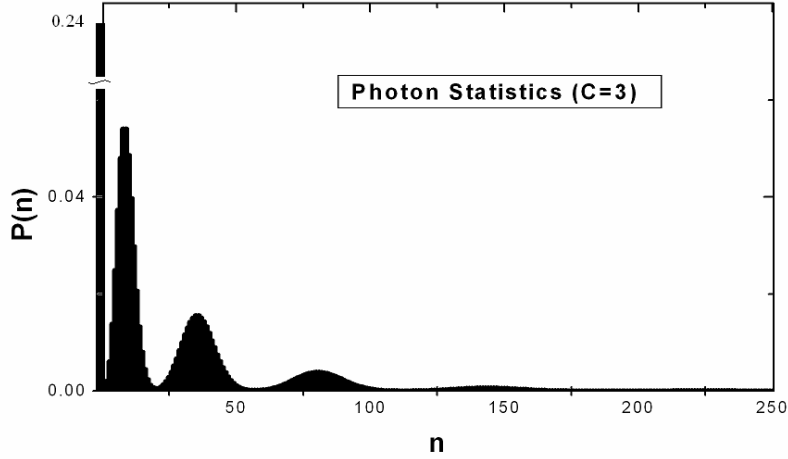}%
\caption{{\footnotesize Numerical plot of the photon number distribution for
}$S=10${\footnotesize \ (20 atoms) and an intentionaly exagerated (for clarity
of ilustration) value of }$C=3${\footnotesize . There are }$11$%
{\footnotesize \ peaks located at }$0,9,36,...,900${\footnotesize \ having
halfwidths }$0,3,6,..30${\footnotesize .}}%
\label{photonstatistics}%
\end{center}
\end{figure}

Figure \ref{photonstatistics} is a plot of this function. Each peak is a
superposition of two equal amplitude coherent states with opposite phase ($\pm
iCM$). The overall envelope is of binomial shape, that reflects the initial
distribution of fluctuations in $S_{z}$. The locations of the peaks are
$0,C^{2},C^{2}2^{2},...C^{2}M^{2},...C^{2}S^{2}$, while the width of an
individual peak radiated by states $\pm M$ is $CM$.%

Although spontaneous emission will limit $C$ to values less than unity, we
consider large values to illustrate the entanglement mechanism. In the limit
$C>>1$, it is seen that the overlap between consecutive peaks decreases since
the separation between them scales as $C^{2}$ while the width only scales as
$C$. The consequence is that a measurement outcome effectively belongs to one
peak only, indicating precisely the collective atomic state onto which the
atoms are projected.

The expectation value and variance of the photon number operator are%

\begin{align}
\left\langle c_{y}^{\dagger}c_{y}\right\rangle  &  =C^{2}\frac{N_{a}}%
{4}\label{av. ph. number}\\
\Delta\left(  c_{y}^{\dagger}c_{y}\right)   &  =C^{2}\sqrt{\frac{N_{a}}%
{4}\{\frac{(N_{a}-1)}{2}+\frac{1}{C^{2}}\}}\nonumber
\end{align}

In view of Eq. (\ref{Climit}) the average number of scattered photons is seen
to be actually limited by $d_{res}$, while a considerable amount of overlap
among different peaks will make the task of separating the atomic states
responsible for scattering very difficult.

\subsection{Null measurement. Spin squeezed states}

From Eq. (\ref{statevectorafterdetection}), the collective atomic state after
a detection event with zero outcome is given by%

\begin{equation}
\left\vert \Psi(T)\right\rangle _{\operatorname{col}}^{n_{m}=0}=\frac
{\overset{S}{\underset{N=-S}{%
{\textstyle\sum}
}}A(S,N)e^{-(CN)^{2}/2}}{\sqrt{\overset{S}{\underset{X=-S}{\sum}}\left\vert
A(S,X)\right\vert ^{2}e^{-(CX)^{2}}}}\left\vert S,N\right\rangle
\end{equation}
%

\begin{figure}
[ptb]
\begin{center}
\includegraphics[
height=3.1038in,
width=3.8104in
]%
{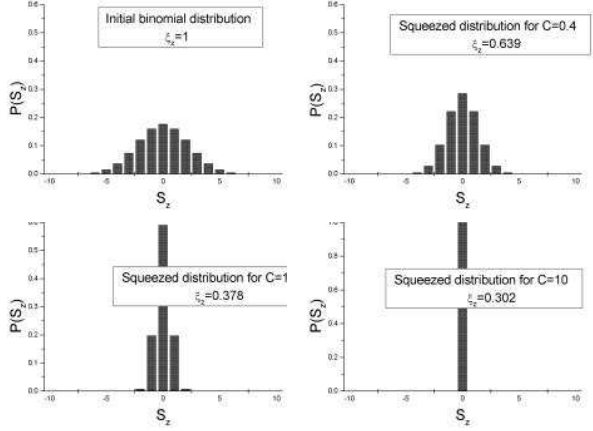}%
\caption{{\footnotesize The initial binomial distribution, for }%
$S=10${\footnotesize , is squeezed owing to the backaction of a measurement
with null outcome. In the absence of decay to transverse modes, an ever better
squeezing parameter is obtained that eventually tends to the Heisenberg limit
when }$C>>1${\footnotesize .}}%
\label{squeezing}%
\end{center}
\end{figure}
\ 

The atomic distribution function given by $P_{a}^{(0)}(M)=\left\vert
A(S,N)\right\vert ^{2}e^{-(CN)^{2}}/g(S,C)$ (where the normalization constant
is given by $g(S,C)=\left[  \overset{S}{\underset{N=-S}{%
{\textstyle\sum}
}}\left\vert A(S,N)\right\vert ^{2}e^{-(CN)^{2}}\right]  ^{1/2}$) is plotted
above showing a squeezing from the initial distribution, owing to the
exponentials in the numerator. An estimate of the width of the squeezed
distribution can be found by using the Stirling approximation for the
factorial of large numbers $N!\simeq\sqrt{2\pi N}\left(  N/e\right)  ^{N}$.
This value $M_{1/e}$ is to be found as the point where the function decreases
to $1/e$ of the initial value $P_{a}(0)=\frac{\left\vert A(S,0)\right\vert
^{2}}{g(S,C)}\simeq\frac{1}{\sqrt{\pi S}g(S,C)}$. Noting that for the initial
distribution $M_{1/e}$ is $\sqrt{S}$ and appreciable decrease of this value is
expected (in the limit $SC^{2}\gg1$), $M_{1/e}$ is found to be given by
$C^{-1}$; remembering that $C$ is proportional to the square root of the total
number of scattered photons (in the forward direction), this is simply stating
that the more photons are incident on the atomic sample, the sharper the
distribution that can be obtained. This result is similar to the conclusions
of Refs. \cite{molmer1, molmer2}.

However, in order to estimate the optimal achievable squeezing, one has to
also analyze the mean spin value. The $y$ and $z$ component expectation values
vanish giving a mean spin pointing in the $x$ direction:%

\begin{equation}
\left\langle S_{z}\right\rangle =0,\ \ \ \ \ \ \left\langle S_{y}\right\rangle
=0,\ \ \ \ \ \ \left\langle \mathbf{S}\right\rangle =\left\langle
S_{x}\right\rangle \widehat{x}\ \ \ \ \ \ \
\end{equation}
Ignoring spontaneous decay, for large values of $C$ this is found to vanish
asymptotically at a rate that is slower than the rate of information gathering
(i.e. the rate of change in the variance of $S_{z}$); the resulting spin
squeezing approaches the Heisenberg limit. This is not true for the realistic
case when decoherence is dominated by the much higher rate [from Eqs. (C12)
and (\ref{Cspon}) it is derived that $C_{spon}^{2}\simeq C^{2}\frac{N_{a}%
}{d_{res}}$]. The expression of the squeezing parameter in the presence of
decay can be thus found (using Eqs.(\ref{sqdef}) and the previous derivations
in the limit $C\gg1/\sqrt{S}$)%

\begin{equation}
\xi\simeq\sqrt{2S}\frac{(\frac{1}{C})}{Se^{-C_{spon}^{2}}}=\frac{\sqrt{2}%
}{\sqrt{S}Ce^{-C^{2}N_{a}/d_{res}}} \label{exactsquezing}%
\end{equation}
The function in the denominator reaches its maximum at $C=\sqrt{d_{res}%
/(2N_{a})}=\frac{1}{2}\sqrt{\frac{d_{res}}{S}}$. This corresponds to a minimum squeezing%

\[
\xi_{\min}=\frac{2\sqrt{e}}{\sqrt{d_{res}}}%
\]
Notice that the best squeezing is obtained when the mean spin reaches a value
around $\sqrt{e}$ smaller than the initial maximal value; this corresponds to
about one photon loss per atom. It should be added, that this being a
probabilistic scheme, its success depends on the likelihood of a null
measurement; this can estimated (for large $C$) as approaching a limiting
value of $\sqrt{\frac{2}{\pi N_{a}}}$ ($\sqrt{\frac{1}{\pi S}}$). It should
also be noted that collective effects have been neglected in estimating the
decay of the mean spin resulting from spontaneous emission. Since one wishes
to work in the limit $n_{a}\lambda^{2}L\gg1$ (the limit usually associated
with superradiance), one should ,in principle, prove that the neglect of
collective effects associated with spontaneous decay is justified.

\subsection{Non-null measurement. "Schrodinger Cat" states}

\begin{figure}
[ptb]
\begin{center}
\includegraphics[
height=3.0805in,
width=3.966in
]%
{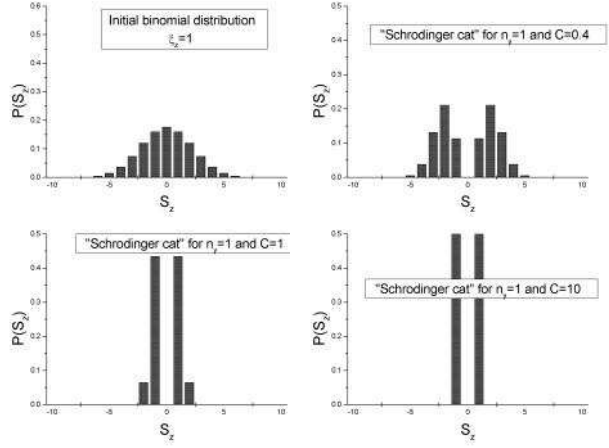}%
\caption{{\footnotesize An outcome of }$1${\footnotesize \ photon projects the
atomic system into a superposition of state which are probable to have given
rise to a }$1${\footnotesize \ photon field. With increasing }$C$%
{\footnotesize \ this superposition sharpens, turning eventually into a pure
Schrodinger cat state.}}%
\label{SCat1}%
\end{center}
\end{figure}

The collapsed atomic state when $n_{m}$ photons are detected is given by%

\begin{equation}
\left\vert \Psi(T)\right\rangle _{\operatorname{col}}=\frac{\overset
{S}{\underset{N=-S}{\sum}}A(S,N)(-iCN)^{n_{m}}e^{-(CN)^{2}/2}}{\sqrt
{\overset{S}{\underset{X=-S}{%
{\textstyle\sum}
}}\left\vert A(S,X)\right\vert ^{2}(CX)^{2n_{m}}e^{-(CX)^{2}}}}\left\vert
S,N\right\rangle
\end{equation}
and gives an atomic probability distribution $P_{a}^{(n_{m})}(M)=\left\vert
A(S,M)\right\vert ^{2}\left(  CM\right)  ^{2n_{m}}e^{-(CN)^{2}}/g(S,C,n_{m})$
shown in Fig \ref{SCat1}.
The localization of the two peaks for given $C,S$ and $n_{m}$ can be made by
finding the values at which the function $\left\vert A(S,M)\right\vert
^{2}\left(  CM\right)  ^{2n_{m}}e^{-(CN)^{2}}$ reaches its maxima. These are
easily found to be given by $\pm M_{m}=\pm\sqrt{n_{m}}/C$. The sharpness of
the peaks is found by using the same procedure as before in the null
measurement case. Denoting the width with $M_{1/e}^{(m)}$, the condition of
half maximum $P_{a}^{(n_{m})}(M_{m}+M_{1/2}^{(m)})=P_{a}^{(n_{m})}(M_{m})/e$
yields the following equation : $\left[  1+M_{1/2}^{(m)}/M_{m}\right]
^{2n_{m}}=e^{C^{2}\left[  2M_{m}M_{1/2}^{(m)}+\left(  M_{1/2}^{(m)}\right)
^{2}\right]  }/e$. A result which proves to set a fundamental distinction
between our scheme and the one presented in Ref. \cite{massar-polzik} is
numerically found, and that is that $M_{1/e}^{(m)}/M_{m}<1$ for any values of
$C$ and $n_{m}$ which indicates that the arms of the cat are always
distinguishable. This is due to the fact that the atomic state corresponding
to a detection of any number of photons other than zero never contains the
Dicke state with zero eigenvalue; therefore a clear separation between the
left and right sides of the atomic population distribution. It should be also
noted that, when $C$ is around a value of $\sqrt{d_{res}/S}$ (which, as seen
in the previous section, leads to optimal squeezing for a null-measurement),
the squeezing parameter for the cat state is approximately equal to $\xi
_{x}\simeq\sqrt{2S}M_{m}/S=\sqrt{n_{m}}/\sqrt{SC^{2}}\simeq\sqrt{n_{m}}%
/\sqrt{d_{res}}$. Eq. (\ref{av. ph. number}) gives an average photon number
(most probable measurement outcome) $\overline{n_{m}}=d_{res}$. The resulting
squeezing parameter $\xi_{x}$ can still be subunitary when the number of
photons detected is smaller than the most probable value $\overline{n_{m}}$.
This result increases the probability to obtain a squeezed state by adding the
category of bimodal spin squeezed states to the single mode obtained with a
zero detection outcome.

\subsection{Subsequent measurements}

\begin{figure}
[ptb]
\begin{center}
\includegraphics[
height=2.4059in,
width=3.96in
]%
{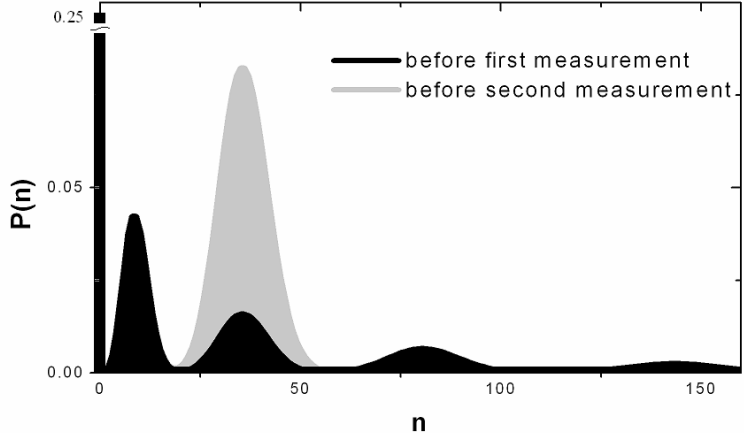}%
\caption{{\footnotesize After the first detection (with }$n_{m}=30$%
{\footnotesize ) projected the atoms onto a Schrodinger cat state with }%
$M=\pm2${\footnotesize , the photon statistics before the second measurement
changes accordingly (the gray plot has only one peak centered around }%
$n=36$){\footnotesize . }}%
\label{SubStat}%
\end{center}
\end{figure}

After running the experiment once and obtaining a collapsed atomic collective
state, this state can be probed by sending in a second pulse and performing
the measurement once again. For well resolved peaks, the photon statistics
before the second measurement contains only the peak that gave the outcome of
the first measurement\ (as shown in Fig. \ref{SubStat}).%

We have performed computer simulations for $N_{a}=20$ atoms. The situation
represented in Fig. \ref{SubStat} describes an ensemble of atoms initially
prepared in a "Schrodinger cat" state. Taking $C=3$ and the measured number of
photons $n_{m}=30$ (which is fairly close to the expectation value of
$C^{2}N_{a}/4=45$ photons and belongs to the second peak, centered at $36$),
after the photodetection process the atomic state is projected onto a "cat"
state with $S_{z}$ projections $\pm2.$ Sending another light pulse through the
ensemble, the expected statistics is modified. A single peak representing a
sum of two coherent states scattered by the states $\pm2$ survives. A next
measurement is most likely only to sharpen the initially prepared "cat" state.

For realistic, small values of $C$, the peaks cannot be well resolved;
therefore, after a single measurement, the resulting cat state will not be
very sharp. However, following the procedure described in Ref.\cite{haroche}
where an initially coherent cavity field is projected onto a Fock state, the
smallness of $C$ can be compensated by continuing to send pulses and detect
the source field until a reasonably good sharpness of the resulting atomic
state is achieved (ideally a projection onto a superposition of Dicke states).

\subsection{Role of detection efficiency}

\begin{figure}
[ptb]
\begin{center}
\includegraphics[
height=2.7224in,
width=3.2534in
]%
{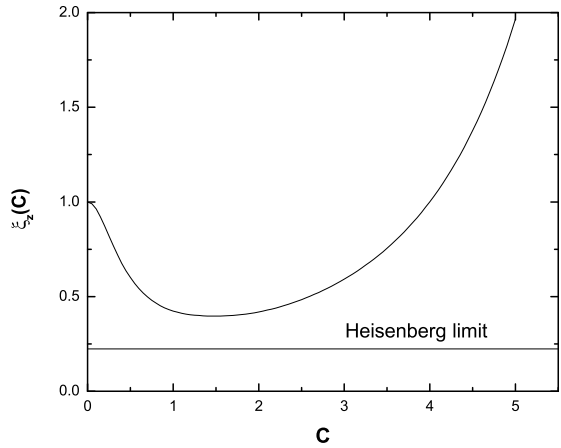}%
\caption{{\footnotesize For }$N_{a}=20${\footnotesize \ and }$85\%$%
{\footnotesize \ detection efficiency, optimal squeezing is reached around
}$C=2${\footnotesize . Thereafter, an increase in the the driving field
strength only deteriorates the squeezing parameter.}}%
\label{inefsq}%
\end{center}
\end{figure}

Less than 100\% efficient detection degrades the quality of squeezing and
robustness of the Schrodinger cat states. Two different approaches have been
used to include the effect of the undetected photons on the collapsed atomic
state. In the first one, one assumes a subunitary efficiency parameter
($\mu<1$) while in the second one, an imaginary beam splitter (with
transmissivity $T$) is placed in the way of the incoming field diverting part
of the light, while the rest undergoes perfect detection ($\mu=1$). Both
calculations yield the same result if $\mu=T$. Using the first approach, the
final state of the total system atoms-scattered field is given by Eq.
(\ref{imperfectdetection}). The non-vacuum field states after detection
represent the part of the field that escaped detection due to inefficiency. A
trace over these states needs to be performed and the density matrix elements
of the final atomic state are given by%

\begin{equation}
\rho_{MN}\left(  \mu\right)  =\frac{A^{\ast}(S,N)A(S,M)\left\{  \alpha
_{N}^{\ast n_{m}}\alpha_{M}^{n_{m}}e^{(1-\mu)\alpha_{N}^{\ast}\alpha_{M}%
}\right\}  e^{-\left(  \left\vert \alpha_{M}\right\vert ^{2}+\left\vert
\alpha_{N}\right\vert ^{2}\right)  /2}}{\overset{S}{\underset{X=-S}{\sum}%
}\left\vert A(S,X)\right\vert ^{2}\left\vert \alpha_{X}\right\vert ^{2n_{m}%
}e^{-\mu\left\vert \alpha_{X}\right\vert ^{2}}}%
\end{equation}
with the notation $\alpha_{M}=-iCM$. For the spin squeezing case, when the
detection outcome is zero, the variance of the population operator is given by%

\begin{equation}
\left(  \Delta S_{z}\right)  ^{2}=\overset{S}{\underset{M=-S}{%
{\textstyle\sum}
}}M^{2}\frac{\left\vert A(S,M)\right\vert ^{2}e^{-\mu\left\vert \alpha
_{M}\right\vert ^{2}}}{\overset{S}{\underset{X=-S}{\sum}}\left\vert
A(S,X)\right\vert ^{2}e^{-\mu\left\vert \alpha_{X}\right\vert ^{2}}}%
\end{equation}
showing a reduction at a slower rate than in the perfect detection case. It
can be shown that, for $C>>1$, the mean spin nevertheless tends to zero at the
same rate as in the perfect detection case. In consequence, the spin squeezing
parameter doesn't tend to the Heisenberg limit with increasing field strength,
but rather reaches a local minimum, tending to infinity afterwards (see
Fig.\ref{inefsq}). The reason is that the decoherence due to the continuous
scattering of photons on the ensemble continues at the same rate no matter
what the detection efficiency is, while the information gathering is slower
for smaller efficiency leading to a slower reduction in $\left(  \Delta
S_{z}\right)  ^{2}$. The value of $C$ at which the squeezing reaches a minimum
is estimated to be in direct relation to the detection efficiency$\;$%

\begin{equation}
C\simeq\frac{1}{\sqrt{1-\mu}}%
\end{equation}%

The Schrodinger cat state with arms at $\pm M_{m}$ is even more sensitive to
the detector inefficiency. Although the height of the two symmetric peaks
[$\rho_{\pm M;\pm N}\left(  \mu\right)  $] is mostly insensitive to the
efficiency, the coherence between them disappears as soon as%

\[
C\gtrsim\frac{1}{M_{m}\sqrt{1-\mu}}%
\]

\section{Conclusions}

A source field approach to the interaction of a Faraday active atomic medium
with an $x$ polarized classical laser pulse has been considered here. The
Faraday effect is responsible for the rotation of the field polarization and
thus the redistribution of initially $x$ polarized photons into the $y$
polarized mode. The photon occupation number of this mode is seen to reflect
the atomic population operator fluctuations; this leads to the observation
that a measurement of the photon number operator provides a means for
information acquisition on the collective atomic quantum state. The key point
in this analysis is that, due to the use of pulsed classical fields rather
than continuous quantized fields, the interaction is not a QND one. This
scheme rather falls into the category of EPR measurements, where the
interaction between subparts of the system has ceased before the actual
measurement takes place. A "discrete" detection formalism is used that allows
to describe the collapse of the system wave function conditioned on an outcome
of $n_{m}$ clicks at the photodetector. This is not much different from the
procedure followed in Ref. \cite{molmer2} where the entanglement of two
macroscopic atomic samples is produced by sending photons one by one through
both media and updating the collective atomic wave function after each detection.

Our results are related to those of Ref. \cite{mabuchi}, when the measurement
time window is matched to our pulse duration $T$. The main difference is that
our choice of ignoring the phase information of the Faraday rotation (through
the use of an unbalanced detection scheme) leads to the preparation of a
superposition of symmetric squeezed states rather than a single squeezed state
shifted to the left or right of the origin.

\section{Acknowledgements}

This work is supported by the National Science Foundation under Grant No.
PHY-0244841 and the FOCUS Center grant.

\section{Appendix A: Maxwell-Bloch approach\qquad}

The incident field is an $x$ polarized laser pulse with a slowly varying
envelope $E_{0}(z,t)$, frequency $\Omega$ and wave number $k_{0}$, propagating
in the $z$ direction. In a circular polarization basis with $\widehat
{\mathbf{\epsilon}}_{+}=-\frac{1}{\sqrt{2}}(\widehat{\mathbf{x}}%
+i\widehat{\mathbf{y}})$ and $\widehat{\mathbf{\epsilon}}_{-}=-\widehat
{\mathbf{\epsilon}}_{+}^{\ast}$, it can be written as a superposition of a
right and left circularly polarized fields%

\begin{equation}
\mathbf{E}(z,t)=-\{\frac{1}{2\sqrt{2}}E_{0}(z,t)e^{ik_{0}z}e^{-i\Omega
t}\widehat{\mathbf{\epsilon}}_{+}+cc\}+\{\frac{1}{2\sqrt{2}}E_{0}%
(z,t)e^{ik_{0}z}e^{-i\Omega t}\widehat{\mathbf{\epsilon}}_{-}+cc\} \tag{A1}%
\end{equation}
We proceed by first finding the phase change in the $\sigma_{+}$ polarized
light, which can induce transitions between $\left\vert g,-1/2\right\rangle $
and $\left\vert e,1/2\right\rangle $.The equations of motion for the density
matrix elements can be solved readily to obtain%

\begin{equation}
\rho_{g,-1/2;e,1/2}(z,t)=\frac{\Lambda(z,t)}{\Delta}\rho_{g,-1/2;g,-1/2}%
(z,t)e^{-i\Omega t} \tag{A2}%
\end{equation}
where
\begin{equation}
\Lambda(z,t)=\frac{1}{2\sqrt{2}\hbar}E_{0}(z,t)e^{ik_{0}z}p \tag{A3}%
\end{equation}
with $p=e\left\langle g,-1/2\left\vert r\right\vert e,1/2\right\rangle $
representing the dipole moment element of the transition. The substitution of
the polarization of the medium

$P(z,t)=np\rho_{g,-1/2;e,1/2}(z,t)+cc$ into the Maxwell-Bloch equation for the
phase shift
\begin{equation}
\left(  \frac{\partial}{\partial z}+\frac{1}{c}\frac{\partial}{\partial
t}\right)  \phi_{+}(z,t)=-\frac{k_{0}}{2\epsilon_{0}}\frac{\operatorname{Re}%
(P(z,t))}{E_{0}(z,t)} \tag{A4}%
\end{equation}
gives, in the steady state regime (assuming that the field depletion is
negligible) a phase shift for the $\sigma_{+}$ polarized light equal to%

\begin{equation}
\phi_{+}=-\frac{np^{2}\Omega L}{\hbar\Delta c\epsilon_{0}}\rho_{g,-1/2;g,-1/2}
\tag{A5}%
\end{equation}
For the $\sigma_{-}$ component of the field, an identical calculation yields

$\phi_{-}=-\frac{np^{2}\Omega L}{\hbar\Delta c\epsilon_{0}}\rho_{g,1/2;g,1/2}%
$. When the atoms are initially only in the two ground sublevels, the Faraday
rotation angle is%

\begin{equation}
\phi=\phi_{+}-\phi_{-}=\frac{np^{2}\Omega L}{\hbar\Delta c\epsilon_{0}}%
[\rho_{g,1/2;g,1/2}-\rho_{g,-1/2;g,-1/2}]=\frac{2p^{2}\Omega}{\hbar\Delta
c\epsilon_{0}A}\left\langle S_{z}\right\rangle \tag{A6}%
\end{equation}

\section{Appendix B: Source-field approach}

Here we look at the same medium as a collection of independent driven atoms
confined in a pencil-shaped volume, that radiate a phased-matched forward
field. The $N$ atoms located at fixed positions $\mathbf{R}_{j}$, are far from
the observer's position $\mathbf{r}$ such that $k\left\vert \mathbf{r}%
-\mathbf{R}_{j}\right\vert \gg1$. The positive frequency field amplitude
operator at position $\mathbf{r}$ is quantized in an infinite volume [with
notation $E_{q}(\omega_{k})=\left(  \frac{\hbar\omega_{k}}{2\varepsilon
_{0}\left(  2\pi\right)  ^{3}}\right)  ^{1/2}$]%

\begin{equation}
\mathbf{E}^{(+)}(\mathbf{r})=i\underset{\lambda}{%
{\textstyle\sum}
}%
{\textstyle\int}
d^{3}kE_{q}(\omega_{k})a_{\lambda}(\mathbf{k})e^{i\mathbf{k}\cdot\mathbf{r}%
}\widehat{\mathbf{\epsilon}}_{\lambda}(\mathbf{k}) \tag{B1}%
\end{equation}
where the continuous field operators obey the usual commutations $\left[
a_{\lambda}^{\dag}(\mathbf{k}),a_{\lambda^{\prime}}(\mathbf{k}^{\prime
})\right]  =\delta(\mathbf{k}-\mathbf{k}^{\prime})\delta_{\lambda
\lambda^{\prime}}$. The atoms in an infinitesimal volume $\delta V(Z)=An\delta
Z$ respond the same way to the external driving field in the approximation
that the transverse profile of the laser beam is constant over the
cross-sectional area $A$. The components of a dimensionless, $Z$ dependent
atomic operator over the $\delta V(Z)$ volume (containing $\delta N_{a}(Z)$
atoms) can be defined as outlined in Ref. \cite{fleischhauer}%

\begin{equation}
\sigma_{\alpha}(Z)=\underset{\delta Z\rightarrow0}{\lim}\frac{1}{\delta
N_{a}(Z)}\underset{Z_{j}\in\delta V(Z)}{%
{\textstyle\sum}
}\sigma_{\alpha}(Z_{j})=\frac{1}{nA}\left(  \underset{\delta Z\rightarrow
0}{\lim}\frac{1}{\delta Z}\underset{Z_{j}\in\delta V(Z)}{%
{\textstyle\sum}
}\sigma_{\alpha}(Z_{j})\right)  \tag{B2}%
\end{equation}
which obey the following commutation relations $\left[  \sigma_{x}%
(Z),\sigma_{y}(Z)\right]  =\frac{2}{nA}\sigma_{z}(Z)\delta(Z-Z^{\prime}).$ The
total spin of the sample with length $L$ is defined as $\mathbf{S}=\frac
{nA}{2}\underset{0}{\overset{L}{%
{\displaystyle\int}
}}\boldsymbol{\sigma}(Z)dZ$ and satisfies angular momentum commutation
relations $\left[  S_{x},S_{y}\right]  =iS_{z}$. The interaction part of the
Hamiltonian (in the Schrodinger picture) can be written%

\begin{align}
H  &  =H_{CF-A}+H_{QF-A}\tag{B3}\\
H_{CF-A}(t)  &  =-\hbar\overset{N_{a}}{\underset{j=1}{%
{\textstyle\sum}
}}\underset{m_{g},m_{e}}{%
{\textstyle\sum}
}\left[  \chi(m_{e},m_{g},z_{j},t)e^{ik_{0}Z_{j}}\sigma_{+}^{(j)}\left(
m_{e},m_{g};Z_{j}\right)  e^{-i\Omega t}+h.c.\right] \nonumber\\
H_{QF-A}  &  =\hbar\overset{N_{a}}{\underset{j=1}{%
{\textstyle\sum}
}}\underset{m_{g},m_{e}}{%
{\textstyle\sum}
}\underset{\lambda}{%
{\textstyle\sum}
}%
{\textstyle\int}
d^{3}k\left[  g_{\lambda}^{\ast}(m_{e},m_{g},\mathbf{k})a_{\lambda}%
(\mathbf{k})\sigma_{+}^{(j)}\left(  m_{e},m_{g};Z_{j}\right)  e^{i\mathbf{k}%
\cdot\mathbf{R}_{j}}+h.c.\right] \nonumber
\end{align}
with the notation $\Delta=\omega-\Omega$, $\Delta_{k}=\omega_{k}-\Omega$,
$\chi(m_{e},m_{g},Z_{j},t)=\frac{E_{0}(Z_{j},t)\left\langle e,m_{e}\left\vert
\widehat{\mathbf{x}}\cdot\mathbf{d}\right\vert g,m_{g}\right\rangle }{2\hbar}%
$(classical field-atoms coupling strength) and $g_{\lambda}(m_{e}%
,m_{g},\mathbf{k})=-\frac{i}{\hbar}E_{q}(\omega_{k})\left\langle
g,m_{g}\left\vert \widehat{\mathbf{\epsilon}}_{\lambda}^{\ast}(\mathbf{k}%
)\cdot\mathbf{d}\right\vert e,m_{e}\right\rangle $ (quantum vacuum-atoms
coupling strength).\ It is convenient now to switch to the Heisenberg picture
and describe the time evolution of time dependent operators as generated by
the Heisenberg picture Hamiltonian. The equation of motion for the
$a_{\lambda}^{HP}(\mathbf{k},t)$, after formal integration, is%

\begin{equation}
a_{\lambda}^{HP}(\mathbf{k,}t)=a_{\lambda}^{HP}(\mathbf{k,}0)e^{-i\omega_{k}%
t}-i\overset{N_{a}}{\underset{j=1}{%
{\textstyle\sum}
}}\underset{m_{g},m_{e}}{%
{\textstyle\sum}
}g_{\lambda}(m_{g},m_{g},\mathbf{k})e^{-i\mathbf{k}\cdot\mathbf{R}_{j}%
}e^{-i\omega_{k}t}\underset{0}{\overset{t}{%
{\textstyle\int}
}}\sigma_{-}^{(j)HP}\left(  m_{e},m_{g};Z_{j},t^{\prime}\right)
e^{i\omega_{k}t^{\prime}}dt^{\prime} \tag{B4}%
\end{equation}
Neglecting the free part (which doesn't contribute to expectation values) and
transforming the summation over atoms into an integral, one can express the
field as%

\begin{align}
\mathbf{E}^{(+)HP}(\mathbf{r},t)  &  =\frac{-i}{\hbar}\underset{\lambda}{%
{\textstyle\sum}
}%
{\textstyle\int}
d^{3}k\overset{N_{a}}{\underset{j=1}{%
{\textstyle\sum}
}}\underset{m_{g},m_{e}}{%
{\textstyle\sum}
}g_{\lambda}(m_{g},m_{g},\mathbf{k})e^{-i\mathbf{k}\cdot\mathbf{R}_{j}%
}e^{-i\omega_{k}t}e^{i\mathbf{k}\cdot\mathbf{r}}\tag{B5}\\
&  \left\{  \underset{0}{\overset{t}{%
{\textstyle\int}
}}\sigma_{-}^{(j)HP}\left(  m_{e},m_{g};Z_{j},t^{\prime}\right)
e^{i\omega_{k}t^{\prime}}dt^{\prime}\right\}  \widehat{\mathbf{\epsilon}%
}_{\lambda}(\mathbf{k})\nonumber
\end{align}
Using the Heisenberg equations of motion one can adiabatically eliminate the
ground-excited level coherences. Owing to the excitation scheme [shown in Fig.
\ref{scheme}(b)], they are connected only to diagonal elements $\sigma
_{g}^{(j)}\left(  m_{g},m_{g};t\right)  $ (ground state populations)%

\begin{equation}
\sigma_{-}^{(j)HP}\left(  m_{e},m_{g};Z_{j},t\right)  =-\frac{\chi(m_{g}%
,m_{e};t)}{\Delta}\sigma_{g}^{(j)}{}^{HP}\left(  m_{g},m_{g};t\right)
e^{ik_{0}Z_{j}}e^{-i\Omega t^{\prime}} \tag{B6}%
\end{equation}
Note that, the population operators do not depend on position (due to the
homogeneity of the medium) or on time (due to the far-off resonance regime in
which we work) and therefore can be replaced by their corresponding operators
in the Schrodinger picture $\sigma_{g}^{(j)}\left(  m_{g},m_{g}\right)  $.
Substituting this result back in Eq. (B5) and summing over polarizations with
$\widehat{\mathbf{\epsilon}}_{1}(\mathbf{k})=\widehat{\mathbf{x}}$ and
$\widehat{\mathbf{\epsilon}}_{2}(\mathbf{k})=\widehat{\mathbf{y}}$, and%
\begin{equation}
\underset{\lambda}{%
{\textstyle\sum}
}g_{\lambda}(m_{g},m_{g},\mathbf{k)}\widehat{\mathbf{\epsilon}}_{\lambda
}(\mathbf{k})=-\frac{i}{\hbar}E_{q}(\omega_{k})\left\langle g,m_{g}\left\vert
\mathbf{d}\right\vert e,m_{e}\right\rangle \tag{B7}%
\end{equation}
we arrive at the electric field operator%

\begin{align}
\mathbf{E}^{(+)}(\mathbf{r},t)  &  =\frac{i}{\hbar\Delta}\underset{m_{g}%
,m_{e}}{%
{\textstyle\sum}
}%
{\textstyle\int}
d^{3}kE_{q}^{2}(\omega_{k})e^{-i\omega_{k}t}e^{i\mathbf{k}\cdot\mathbf{r}%
}\tag{B8}\\
&  \left\{  \underset{0}{\overset{t}{%
{\textstyle\int}
}}dt^{\prime}\chi(m_{g},m_{e};t^{\prime})\left[  \overset{N_{a}}%
{\underset{j=1}{%
{\textstyle\sum}
}}e^{-i\left(  k_{x}X_{j}+k_{y}Y_{j}\right)  }e^{i(k_{0}-k_{z})Z_{j}}%
\sigma_{g}^{(j)}\left(  m_{g},m_{g}\right)  \right]  e^{i\omega_{k}t^{\prime}%
}e^{-i\Omega t^{\prime}}\right\}  \left\langle g,m_{g}\left\vert
\mathbf{d}\right\vert e,m_{e}\right\rangle \nonumber
\end{align}
The sum over the atoms can be transformed into an integral. In the limit of
large transverse area $A\gg\lambda^{2}$, the integral over $X$ and $Y$ is
performed, yielding%

\begin{equation}
\overset{N_{a}}{\underset{j=1}{%
{\textstyle\sum}
}}e^{-i\left(  k_{x}X_{j}+k_{y}Y_{j}\right)  }e^{i(k_{0}-k_{z})Z_{j}}%
\sigma_{g}^{(j)}\left(  m_{g},m_{g}\right)  =4\pi^{2}n\delta(k_{x}%
)\delta(k_{y})\overset{L}{\underset{0}{%
{\textstyle\int}
}}dZe^{i(k_{0}-k_{z})Z}\sigma_{g}\left(  m_{g},m_{g}\right)  \tag{B9}%
\end{equation}
Grouping all the terms that depend on $\omega_{k}$, and evaluating the slowly
varying term $E_{q}^{2}(\omega_{k})$ at $\Omega$, the integral over the field
modes becomes $%
{\textstyle\int}
d\omega_{k}e^{-i\omega_{k}(t-t^{\prime}-z/c+Z/c)}=$ $\delta(t-t^{\prime
}-z/c+Z/c)$. Evaluating the $t^{\prime}$ integral now by replacing $t^{\prime
}=t-z/c+Z/c$, the $e^{ik_{0}Z}$ term is canceled, which allows us to perform
the integral over $Z$ and introduce collective population operators
$S_{g}\left(  m_{g},m_{g}\right)  =\frac{nA}{2}\underset{0}{\overset{L}{%
{\displaystyle\int}
}}dZ\sigma_{g}^{(j)}\left(  m_{g},m_{g}\right)  $. The total field becomes%

\begin{align}
\mathbf{E}^{(+)HP}(\mathbf{r},t)  &  =\frac{8i\pi^{2}E_{q}^{2}(\Omega)}%
{\hbar\Delta cA}\left\{  \underset{m_{g},m_{e}}{%
{\textstyle\sum}
}\chi(m_{g},m_{e};t)S_{g}\left(  m_{g},m_{g}\right)  \left\langle
g,m_{g}\left\vert \mathbf{d}\right\vert e,m_{e}\right\rangle \right\}
\tag{B10}\\
&  e^{-i\Omega(t-z/c)}\nonumber
\end{align}
Separating the field into $x$ and $y$ components, and noting that
$S_{g}(1/2,1/2)+S_{g}(-1/2,-1/2)=S$ and $S_{g}(1/2,1/2)-S_{g}(-1/2,-1/2)=S_{z}%
$ one obtains the following expressions for the field at position $z$%

\begin{align}
E_{x}^{(+)HP}(z,t)  &  =\frac{-inL\Omega\left\vert p\right\vert ^{2}}%
{2\hbar\Delta\varepsilon_{0}c}E_{0}(0,t-z/c)e^{-i\Omega(t-z/c)}\tag{B11}\\
E_{y}^{(+)HP}(z,t)  &  =\left\{  \frac{nL\Omega\left\vert p\right\vert ^{2}%
}{2\hbar\Delta\varepsilon_{0}cS}E_{0}(0,t-z/c)e^{-i\Omega(t-z/c)}\right\}
S_{z}\nonumber
\end{align}
A Faraday rotation operator can be defined as:%

\begin{equation}
\widehat{\phi}=2E_{y}^{(+)}(z,t)e^{i\Omega(t-z/c)}/E_{0}(z/c,t)=\frac
{nL\Omega\left\vert p\right\vert ^{2}}{\hbar\Delta\varepsilon_{0}c}\frac
{S_{z}}{S} \tag{B12}%
\end{equation}
whose expectation value coincides with the result obtained using a
Maxwell-Bloch approach.

\section{Appendix C: Derivation of the effective Hamiltonian}

An effective Hamiltonian that describes the generation of a quantized source
field by the driven atomic system can be derived from $H_{QF-A}$ in Eq. (B3).
The steps that are necessary in this procedure are discussed here. First,
using the previously defined continuous atomic operators, one can transform
the sum over atoms into an integral $\overset{N_{a}}{\underset{j=1}{%
{\textstyle\sum}
}}\sigma_{\alpha}^{j}e^{i\mathbf{k}\cdot\mathbf{R}_{j}}\rightarrow n_{a}%
{\displaystyle\int}
dXdYdZ\sigma(Z)e^{i\mathbf{k}\cdot\mathbf{R}}=n_{a}\overset{L}{\underset{0}{%
{\textstyle\int}
}}dZ\sigma(Z)\underset{A}{%
{\textstyle\int}
}dXdYe^{i\mathbf{k}\cdot\mathbf{R}}$ . An integration over the $x$ and $\ y$
components of $\mathbf{k}$ followed by one over $X$ and $Y$ allows us to
define an effective one-mode continuous field operator that describes photons
propagating in the $z$ direction with polarization $\lambda$ and transverse
spatial extend $A$
\begin{equation}
d_{\lambda}(k_{z})=\frac{1}{2\pi\sqrt{A}}\underset{A}{%
{\textstyle\int}
}dXdY%
{\textstyle\int}
dk_{x}dk_{y}a_{\lambda}(\mathbf{k})e^{ik_{x}X}e^{ik_{y}Y} \tag{C1}%
\end{equation}
The commutation relations are $\left[  d_{\lambda}(k_{z}),d_{\lambda^{\prime}%
}^{\dag}(k_{z}^{\prime})\right]  =\delta(k_{z}-k_{z}^{\prime})\delta
_{\lambda\lambda^{\prime}}$\ With the newly defined field operator and
continuous atomic operators replaced in Eq. (B3) the expression for the
interaction part of the Hamiltonian becomes
\begin{equation}
H_{QF-A}=2\pi\hbar nA^{1/2}\underset{m_{g},m_{e},\lambda}{%
{\displaystyle\sum}
}%
{\textstyle\int}
dk_{z}\overset{L}{\underset{0}{%
{\textstyle\int}
}}dZ\left\{  g_{\lambda}(m_{g},m_{e};k_{z})d_{\lambda}^{\dag}(k_{z})\sigma
_{-}\left(  m_{e},m_{g};Z\right)  e^{-ik_{z}Z}+h.c\right\}  \tag{C2}%
\end{equation}
Using the equivalent of Eq. (B6) for continuous operators, one can replace
coherences by population operators (in the Heisenberg picture)%

\begin{equation}
\sigma_{-}^{HP}\left(  m_{e},m_{g};Z,t\right)  =-\frac{\chi(m_{g},m_{e}%
;t)}{\Delta}\sigma_{g}^{HP}\left(  m_{g},m_{g};t\right)  e^{ik_{z}%
Z}e^{-i\Omega t} \tag{C3}%
\end{equation}

The Heisenberg picture Hamiltonian is found in terms of population operators.
Negligibly small Langevin fluctuations have been ignored here since their
correlations are smaller than the term driving the coherence by a factor of
$\gamma/\Delta$. With a transformation back to the Schrodinger picture, an
expression for the effective Hamiltonian is found. With removal of the free
evolution of field and atoms, and neglecting the light shift of the lower
levels, an interaction picture Hamiltonian that will govern the evolution of
the system is found to be%

\begin{equation}
H_{QF-A}^{IP}=2\pi\hbar nA^{1/2}\underset{m_{g},m_{e},\lambda}{%
{\displaystyle\sum}
}%
{\textstyle\int}
dk_{z}\overset{L}{\underset{0}{%
{\textstyle\int}
}}dZ\left\{
\begin{array}
[c]{c}%
\frac{g_{\lambda}(m_{g},m_{e};k_{z})\chi(m_{g},m_{e};t)}{\Delta}d_{\lambda
}^{\dag}(k_{z})\\
\sigma_{g}\left(  m_{g},m_{g}\right)  e^{i(k_{0}-k_{z})Z}e^{i(\omega
_{k}-\Omega)t}+h.c
\end{array}
\right\}  \tag{C4}%
\end{equation}

Since the regime we work in is the low saturation limit where the absorption
of the field is negligible, the population operators are not changed by the
spatial dependence of the field amplitude and therefore $Z$ independent.
Noting that the integral over $k_{z}$ varies rapidly outside a small interval
around $k_{0}$, $g_{\lambda}(m_{g},m_{e};k_{z})$ (which varies slowly as
$\sqrt{k_{z}}$) can be evaluated at $k_{0}.$ The time dependence of the
coupling strength includes $E(0,t)$ which is the slowly varying envelope of
the field with Fourier components $E(\omega_{k}-\Omega)$ that are nonzero only
in a small interval centered at the carrier frequency $\left(  \omega
_{k}-\Omega\right)  \ll1/T$. Consequently, the allowed interval for the wave
vectors is $\left(  k_{0}-k_{z}\right)  \ll1/cT$. In the limit of $L\ll cT$
(all atoms see the same field amplitude at one instant in time), it is implied
that $\left(  k_{0}-k_{z}\right)  Z\ll Z/cT<L/cT\ll1$. This condition allows
us to set the spatial dependence $e^{i(k_{0}-k_{z})Z}$ to $1$ and evaluate
$\chi(m_{g},m_{e},Z,t)$ at $Z=0$.\ The integral over the sample length
introduces the collective spin operator. That means that the forward scattered
field couples only to the symmetric atomic mode through the lowering and
raising spin operators $S_{+},S_{-}$. The sum over $m_{g},m_{e}$ can be
performed now for each polarization ($x$ and $y$) with the result for the
simplified Hamiltonian:%
\begin{align}
H_{QF-A}^{IP}  &  =\left\{  \frac{2i\pi\left\vert p\right\vert ^{2}}%
{\hbar\Delta\sqrt{A}}E_{q}(\Omega)E(0,t)\left(
{\textstyle\int}
dk_{z}d_{x}^{\dag}(k_{z})e^{i(\omega_{k}-\Omega)t}\right)  S+h.c\right\}
+\tag{C5}\\
&  \left\{  \frac{2\pi\left\vert p\right\vert ^{2}}{\hbar\Delta\sqrt{A}}%
E_{q}(\Omega)E(0,t)\left(
{\textstyle\int}
dk_{z}d_{y}^{\dag}(k_{z})e^{i(\omega_{k}-\Omega)t}\right)  S_{z}+h.c\right\}
\nonumber
\end{align}
Notice that the first term in the rhs commutes with both $y$ mode field
operators and atomic operators, which means that the $x$ polarized field is
decoupled from the rest of the system. Equivalently, one can say that only the
$y$ part carries information about the quantum state of the atomic ensemble.
The $x$ part can thus be discarded, and setting $B=\frac{2\pi\left\vert
p\right\vert ^{2}}{\hbar^{2}\Delta\sqrt{A}}E_{q}(\Omega)$ the $y$ part takes
the form:%

\begin{equation}
\left(  H_{QF-A}^{IP}\right)  _{y}=\hbar BE(0,t)\left(
{\textstyle\int}
dk_{z}d_{y}^{\dag}(k_{z})e^{i(\omega_{k}-\Omega)t}\right)  S_{z}+h.c \tag{C6}%
\end{equation}
With the observation that
\begin{equation}
\left[  \left(  H_{QF-A}^{IP}\right)  _{y}(t),\left(  H_{QF-A}^{IP}\right)
_{y}(t^{\prime})\right]  =0 \tag{C7}%
\end{equation}
the evolution operator can be written in a simple form:%

\begin{equation}
U(T)=\exp\left[  -\frac{i}{\hbar}\overset{T}{\underset{0}{%
{\textstyle\int}
}}\left(  H_{QF-A}^{IP}\right)  _{y}(t)dt\right]  \tag{C8}%
\end{equation}
The time integral brings the Fourier components of the incident pulse field
envelope $\underset{0}{\overset{T}{%
{\textstyle\int}
}}dte^{i(\omega_{k}-\Omega)t}E(0,t)\simeq E(\omega_{k}-\Omega)$ giving%

\begin{equation}
U(T)=\exp\left[  -i\left(  B\left\{
{\textstyle\int}
dk_{z}E(\omega_{k}-\Omega)d_{y}^{\dag}(k_{z})\right\}  S_{z}+h.c\right)
\right]  \tag{C9}%
\end{equation}
The integral over the continuous creation operators can be represented by an
effective one-photon creation operator with carrier frequency $\Omega$ and
duration $\left(  c\Delta k\right)  ^{-1}\simeq T$ and obeying the commutation
relation $[c_{y},c_{y}^{\dag}]=1$%

\begin{equation}
c_{y}^{\dag}=\frac{c^{1/2}}{\sqrt{\overset{T}{\underset{0}{%
{\textstyle\int}
}}dt\left\vert E(t)\right\vert ^{2}}}%
{\textstyle\int}
dk_{z}E(\omega_{k}-\Omega)d_{y}^{\dag}(k_{z}) \tag{C10}%
\end{equation}
which leads to a simple form for the evolution operator%

\begin{equation}
U(T)=\exp[-iC(c_{y}^{\dag}-c_{y})S_{z}] \tag{C11}%
\end{equation}
with $C=\frac{B}{c^{1/2}}\sqrt{%
{\textstyle\int}
dk_{z}\left\vert E(\omega_{k}-\Omega)\right\vert ^{2}}=\frac{2\pi\left\vert
p\right\vert ^{2}E_{q}(\Omega)}{\hbar^{2}\Delta\sqrt{A}c^{1/2}}\sqrt
{\overset{T}{\underset{0}{%
{\textstyle\int}
}}dt\left\vert E(t)\right\vert ^{2}}.$ In terms of one atom total spontaneous
emission loss during the time $T$ of the applied laser pulse [see Eq.
\ref{Cspon}] $C$ can also be expressed as
\begin{equation}
C=\left[  \frac{3}{16\pi^{2}}\left(  \frac{\lambda^{2}}{A}\right)  \left(
\gamma\overset{T}{\underset{0}{%
{\textstyle\int}
}}dt\frac{\left\vert \chi(t)\right\vert ^{2}}{\Delta^{2}}\right)  \right]
^{1/2}=\left[  \frac{3}{16\pi^{2}}\left(  \frac{\lambda^{2}}{A}\right)
C_{spon}\right]  ^{1/2} \tag{C12}%
\end{equation}

\bigskip

\bigskip\

\end{document}